\documentclass[twocolumn,prl,floatfix,showpacs]{revtex4}

\usepackage{bm}
\usepackage{amsmath,amssymb}
 \usepackage{palatino}
 \usepackage[osf]{mathpazo}
 \usepackage[dvips]{graphicx}
 \usepackage[dvips]{color}
 \usepackage{pslatex}

\def\matr#1{{\ensuremath{\underline{\underline{ {\bm{#1}} }}}}}

\def\vec#1{{\ensuremath{\bm{#1}}}}

\def\half{{\textstyle \frac{1}{2}}}

\def\third{{\textstyle \frac{1}{3}}}

\def\n{ {\vec{n} }}

\def\rvec{\vec{r}}

\def\d{\textrm{d}}

\def\R{{\vec{R}}}

\def\s#1{_{\rm #1} }
\def\sp#1{^{\rm #1} }

\def\lm{\matr{\lambda} }

\def\Det#1{{\rm Det}\left( #1 \right)}

\def\ber{\begin{eqnarray}}
\def\eer{\end{eqnarray}}
\def\be{\begin{equation}}
\def\ee{\end{equation}}
\def\bea{\begin{eqnarray}}
\def\eea{\end{eqnarray}}

 \def\ugamma{{\underline \gamma }}

\begin{document}

\title{Negative Gaussian curvature from induced metric changes}
\author{Carl D. Modes$^1$ and Mark Warner$^2$}
 \affiliation{$^{1}$ The Rockefeller University, Center for Studies in Physics and Biology, 1230 York Ave, New York, NY 10065, USA\\$^{2}$ Cavendish Laboratory, University of Cambridge, 19 JJ Thomson Avenue, Cambridge CB3 0HE, UK}
\date{\today}

\begin{abstract}
We revisit the light or heat-induced changes in topography of initially flat sheets of solid that elongate or contract along patterned, in-plane director fields.  For radial or azimuthal directors, negative Gaussian curvature is generated -- so-called ``anti-cones".  We show that azimuthal material displacements are required for the distorted state to be stretch-free and bend-minimising.  The resultant shapes are smooth and aster-like and can  become re-entrant in the azimuthal coordinate for large deformations. We show that care is needed when considering elastomers rather than glasses, though the former offer huge deformations.

\end{abstract}
 \maketitle

Differential growth in planar sheets \cite{Ben-Amar:08a,Ben-Amar:08b} induces topographical change in order to avoid or minimize stretch energy.  Analogous, solid-nematic sheets are particularly rich since light or heat-induced length changes \cite{Mol:05,Mol:91,Finkphoto,Harris:05,vanoosten:07} are easily and reversibly driven.  Further, their director fields $\n(\rvec)$ that give the principal direction of elongation or contraction can be written in order to obtain any incompatibility of in-plane deformation that is then resolved by topographical change and out-of-plane buckling, resulting in a desired Gaussian curvature.

We revisit the simplest systems generating negative Gaussian curvature, namely anti-cones \cite{ModesPRE:10,ModesPRS:11} (observed by Broer \textit{et al} \cite{HaanAC:12}) that are the synthetic analogs of the ruffles, or asters, of Ben Amar \textit{et al} \cite{Ben-Amar:08a,Ben-Amar:08b}. In particular we examine pathologies arising in some modelling because of restrictions unnecessarily placed on shell deformations, and we present here bend-minimising solutions that are the optimal choice in the manifold of zero-stretch topographical changes from localised negative Gaussian curvature.

Effectively, heated or illuminated sheets of nematic solid suffer changes of metric and hence changes in Gaussian curvature and thus of shape \cite{ModesPRE:11,ModesEPL:12}.  This metric change approach also has recently been exploited in such problems in an ambitious work also concerned with general shape determination \cite{AharoniPRL:14}.
Our analysis is equivalent, dealing with the differential geometry of space curves generating these surfaces.  We re-examine small amplitude, stretch-avoiding solutions, since they point to differences with recent work \cite{PismenPRE:14}, before we present large amplitude, stretch-free, bend-minimising results. We will also discuss the role of nematic elastomers (rather than glasses) that have mobile directors and can respond more subtly to Gaussian curvature change.

\begin{figure}[!b]
\centerline{\includegraphics[width=8.5cm]{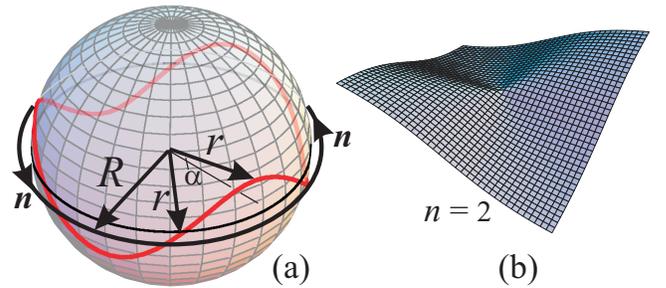}}
\caption{(Color online) (a) An initially flat disc of radius $R$ deformed into an anti-cone with an in-material radius $r$. Director lines are circular, see for instance an $\n(\rvec)$ indicated at radius $R$. (b) an $n=2$ ``anti-cone".}
\label{fig:anticones}
\end{figure}

Consider concentric circles of $\n$ in a flat sheet of nematic glass initially occupying the equatorial plan of the sphere in Fig.~\ref{fig:anticones}(a). Let the elongation along $\n(\rvec)$ be by a factor of $\lambda > 1$ on cooling or return to the dark (or, equivalently, radially arrayed $\n$ subjected to heating or light exposure with $\lambda < 1$). There is a corresponding contraction by $\lambda^{-\nu}$ in the two perpendicular directions, one in-plane and the other through the thickness.  Volume changes locally by a factor of $\Det{\lm} = \lambda^{1-2\nu}$ where $\lambda$, $\lambda^{-\nu}$, $\lambda^{-\nu}$ are the elements of the deformation gradient tensor $\lm$ in its principal frame (based upon $\n$).  The opto-thermal Poisson ratio $\nu$ takes values typically in the range $(\third,2)$ in nematic glasses  \cite{Harris:05}, and $\nu = \half$ in nematic elastomers where volume is conserved.  Note that area would be conserved for $\nu = 1$, as assumed by Pismen \cite{PismenPRE:14}, but this value is of \textit{no} particular physical significance. If elastic stretches are to be avoided on cooling/darkness, then one simply requires that, in deformation to a different topography, the deformation gradients take principal values $\lambda$ and $\lambda^{-\nu}$ corresponding to the natural opto-thermal value under those conditions.
We thereby automatically avoid stretch energy costs.

In cylindrical coordinates, a reference state point is $\R = (R,\Phi,Z=0)$ in the initially flat disc, the equatorial plane of the sphere in Fig.~\ref{fig:anticones}(a).  On cooling, its image in the target state is $\rvec = (\rho, \phi,z=h(\phi))$, where $h$ is the elevation from the initial plane (note the use of lower and upper case variables). The target state curve is on a sphere of
radius $r$, see the trajectory in  Fig.~\ref{fig:anticones}(a), and the in-material radius is now
\be
r^2 = \rho^2 + z^2 \;\; \rightarrow r = \rho \sqrt{1 + (h/\rho)^2}\label{eq:rad}.
\ee
We take a form for distortion that costs no stretch energy and will also minimise bend energy for small amplitudes:
\be
h = \rho A \sin(n\phi) \label{eq:hdef},
\ee
with integer $n$  for closure.
The amplitude $A$, after scaling by the cylindrical radius $\rho$, is $A = \tan\alpha$; see Fig.~\ref{fig:anticones}(b). The angle $\alpha$ is made between an anti-cone generator at a displacement antinode and the equatorial plane initially taken by the flat disc; see Fig.~\ref{fig:anticones}(a).  The amplitude needed to take up the extra length of the perimeter with respect to the radius is a global requirement \cite{ModesPRE:10}, not local, that we also return to.

An element $\d s$ of length in the target space image of the reference space element $R \d \Phi$ is
\be
\d s = \d\phi \sqrt{ \left(\frac{\partial\rho}{\partial \phi} \right)^2 +
\rho^2 + \left(\frac{\partial h}{\partial\phi} \right)^2}\label{eq:diff-perimeter}.
\ee
We now require the sheet deforms locally according to $\lm$:
\be
r  = \lambda^{-\nu} R \;\; \textrm{and} \;\; \d s = \lambda (R \,\,\d\Phi) \rightarrow \frac{\partial s}{R\,\,\partial\Phi} = \lambda \label{eq:transforms}.
\ee
From eqn.~(\ref{eq:rad}) we have
\bea
\rho(\phi) &=& \lambda^{-\nu} R/\left( 1 + A^2 \sin^2 n \phi \right)^{\frac{1}{2}}\label{eq:rho}\\  h(\phi) &=& \lambda^{-\nu} R \, A \sin(n\phi)/\left( 1 + A^2 \sin^2 n \phi \right)^{\frac{1}{2}} \label{eq:h}.
\eea
Returning relations~(\ref{eq:rho}) and (\ref{eq:h}) to eqn~(\ref{eq:diff-perimeter}) yields
\be
\d s  = \d\phi \,\,\lambda^{-\nu} R\frac{\left[1 + A^2 +(n^2-1)A^2\cos^2 n \phi\right]^{1/2} }{\left( 1 + A^2 \sin^2 n \phi \right)}\label{eq:new-dp}.
\ee
Dividing Eqn.~(\ref{eq:new-dp}) by $R\d\Phi$, using eqn.~(\ref{eq:transforms}), and rearranging gives
\be
\frac{\partial\phi}{\partial\Phi}  = \lambda^{1+\nu} \frac{\left( 1 + A^2 \sin^2 n \phi \right)}{\left[1 + A^2 +(n^2-1)A^2\cos^2 n \phi \right]^{1/2}}\label{eq:phi-var}.
\ee
The image's azimuthal angle $\phi$ differs from the $\Phi$ of the original target state point.
Respect for the natural distortions of the sheet has been built in.  There are no stretch/compression costs, but some bend energy that we return to.

\begin{figure}[]
\centerline{\includegraphics[width=7cm]{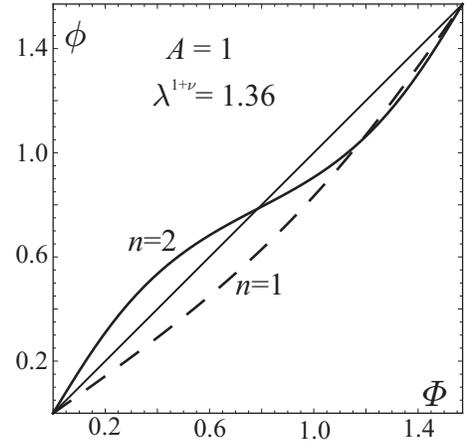}}
\caption{Variation of the image azimuthal angle $\phi$ with its reference $\Phi$ for distortion to an anti-cone for $n=2$, using the infinitesimal form~(\ref{eq:hdef}), but for clarity with reasonably large distortion $\lambda^{1+ \nu} = 1.36$, implying an amplitude $A=1$ of the aster-like structure.  The trivial case $n=1$ of the body rotation of a disc has $\lambda=1$, no distortion, but the same rotation ($\pi/4$, about a diameter) is taken as that suffered by the antinodal generators in the $n=2$ illustration.}
\label{fig:phi_var}
\end{figure}

Pismen \cite{PismenPRE:14} restricts all distortions in his model of anti-cones to be meridional and radial as in the more straightforward case of simple cones, whereas these distortions to anti-cones described above suffer azimuthal displacements too. The differential method employed above is in effect equivalent to the metric method of Pismen, his eqns.~(12)--(14), but with differing assumptions (about $\Phi$ and $\phi$) with the result that our anti-cones, eq.~(\ref{eq:hdef}), do not have creases.
It is interesting to integrate relation~(\ref{eq:phi-var}) to give $\phi(\Phi)$, see Fig.~\ref{fig:phi_var}, to see for instance the azimuthal variation in this model for an $n=2$ anti-cone which repeats at $\Phi = \phi = \pi$.  It has its first node at $\Phi = \phi = \pi/2$, and its first antinode at $\Phi = \phi = \pi/4$.  At both these points it is clear, for symmetry reasons, that $\Phi = \phi$. The deviation of $\phi$ from $\Phi$ is the essential difference between our results and those of Pismen, and is the mechanism by which creases are here avoided.

The case $n=1$ is simply a uniform body rotation by an $\alpha$ about a diameter since all radii at $\phi$ and $\phi + \pi$ are simple continuations of each other and the disc remains flat.  Clearly $\lambda =1$ for body rotations. For $n \ge 2$ the rotation by $\alpha$ such that $A= \tan\alpha$ is identifiable only at the antinodal lines. Though trivial, the $n=1$ case is instructive: simple geometry applied to the $n=1$ transformation gives the mapping of the azimuthal angle:
\be
\tan\phi = \cos\alpha \tan\Phi \label{eq:n1-phi}.
\ee
Alternatively, explicit integration of eqn.~(\ref{eq:phi-var}) is trivial and yields (\ref{eq:n1-phi}) for $\lambda = 1$ and useing $\sqrt{1+A^2} = \sec\alpha$.  This analysis of simple rotation is a motivation for adopting $h=\rho A \sin(n\phi)$ for the axial distortions into an anti-cone, and underscores a need to have a $\phi(\Phi)$ for non trivial cases when it already arises for $n=1$.

The method of \cite{ModesPRE:10,ModesPRS:11} for anti-cones was to take a global version of Eqn.~(\ref{eq:new-dp}) by integrating over $s$ to give the whole new perimeter $C$, the surplus of which relative to the radius determines the amplitude as a function of length changes, $A(\lambda)$.  Since $\n$ is along the perimeter, $C = \lambda 2\pi  R$.  The integral of the right hand side of Eqn.~(\ref{eq:new-dp}) over $\phi = (0,2\pi)$ yields $\lambda^{-\nu} 2 \pi I(A,n)$ where the integral $I$ is eqn~(6) of \cite{ModesPRE:10}:
  \be I(n,A) = \int_0^1
du \sqrt{\frac{n^2 A^2 \cos^2 2 \pi n u }{(1+A^2 \sin^2 2 \pi n
u)^2} + \frac{1}{1+A^2 \sin^2 2 \pi n u}}\;.\label{eq:phi-length}\ee 
One sees from the current analysis that a stretch-free state is guaranteed globally since it is built-in locally.  However one cannot fully specify this problem locally.  From the above, the amplitude $A$ and the distortion $\lambda$ are connected through $I(A,n) = \lambda^{1+\nu}$.  As $\lambda$ changes, so too must the amplitude $A$ of the ruffles in order that all surplus length around a perimeter is accommodated.  The negative  Gaussian curvature localized
at the apex of the anti-cone is $2\pi(1-I)$.  Eventual re-entrance and the transition to higher $n$ anti-cones are discussed in \cite{ModesPRE:10,ModesPRS:11} and below.


\noindent \textit{Finite deformations}:  Denote by $\ugamma(s)$ a position on a curve on the deformed surface with unit distance $r=1$ from the origin; see Fig.~\ref{fig:anticones}(a). It will have evolved from the circle with $R = r\lambda^{\nu}=\lambda^{\nu}$ in the initial flat disc since these radii are perpendicular to $\n$ and contract by a factor of $\lambda^{-\nu}$.  The initial perimeter is therefore $2\pi \lambda^{\nu}$ which then changes by a factor of $\lambda$ (since it is along $\n$) to $2\pi \lambda^{1+ \nu}$.  Stretch is avoided and it remains to minimize bend energy.

In the Darboux frame of the curve $\ugamma(s)$ with respect to the deformed surface one has
\bea
\vec{T}(s) &=& \ugamma'(s)\; ; \;\;\; |\ugamma'| = 1\;\;\;\; \textrm{tangent} \nonumber\\
\vec{u}(s) &\equiv& \vec{u}(\ugamma(s))\;\;\;\;\;\;\;\;\;\;\;\;\;\;\;\;\;\; \textrm{unit surface normal}\nonumber\\
\vec{t}(s) &=& \vec{u}(s) \wedge \vec{T}(s)\;\;\;\;\;\;\;\;\;\; \textrm{tangent normal}
\eea
 with $'$ denoting $\d/\d s$. The rate of change of the tangent gives the two components of bend:
 \be
 \vec{T}'(s) = \kappa\s{n} \vec{u}(s) + \kappa\s{g}\vec{t}(s) \label{eq:bends}
 \ee
where $\kappa\s{n}=\vec{u}(s)\cdot \vec{T}'$ and $\kappa\s{g}=\vec{t}(s)\cdot \vec{T}'$ are respectively the normal and geodesic curvatures.  We need to minimize the integral of $\kappa\s{n}^2$ over the deformed surface.

The surface is described by the scale-invariant curves $\rvec(s) = r \ugamma(s)$.  By construction $\ugamma(s)\cdot\ugamma(s) = 1$ whence $\ugamma'(s)\cdot\ugamma(s) \equiv \ugamma'\cdot \hat{\rvec} = 0$.  For clarity later, we have written $\ugamma = \hat{\rvec}$, a unit vector in the surface to the curve.  It is also perpendicular to $\ugamma'$ and thus is $\hat{\rvec} = \vec{t}$.  From the Darboux triad we have $\vec{u} = \hat{\rvec} \wedge \ugamma'$, whence from eq.~(\ref{eq:bends}) we have
\be
\kappa\s{n} = (\hat{\rvec}\wedge \ugamma')\cdot \ugamma''\label{eq:normalbend}.
\ee
The total bend energy is proportional to
\be
\int \d r . r\d s \left((\hat{\rvec}\wedge \ugamma')\cdot \frac{\ugamma''}{r}\right)^2 = \int \frac{\d r}{r}\d s \left((\hat{\rvec}\wedge \ugamma')\cdot \ugamma''\right)^2
\label{eq:normalbend_energy}.
\ee
where $\int\d s$ is over $s \in (0, 2\pi\lambda^{1+\nu})$. The $1/r$ with the $\ugamma''$ arises because for $r \ne 1$, the derivative $\vec{T}'$ is really $\frac{1}{r} \frac{\d \vec{T}}{\d s}= \frac{\ugamma''}{r}$.  The high bend energy density at the apex of the anti-cone is discussed in quantitative detail in \cite{ModesPRS:11}, being smoothed out by some stretch and in any event probably not arising because of director escape into the third dimension during fabrication.

\begin{figure}
\centering
\includegraphics[width=1.4in]{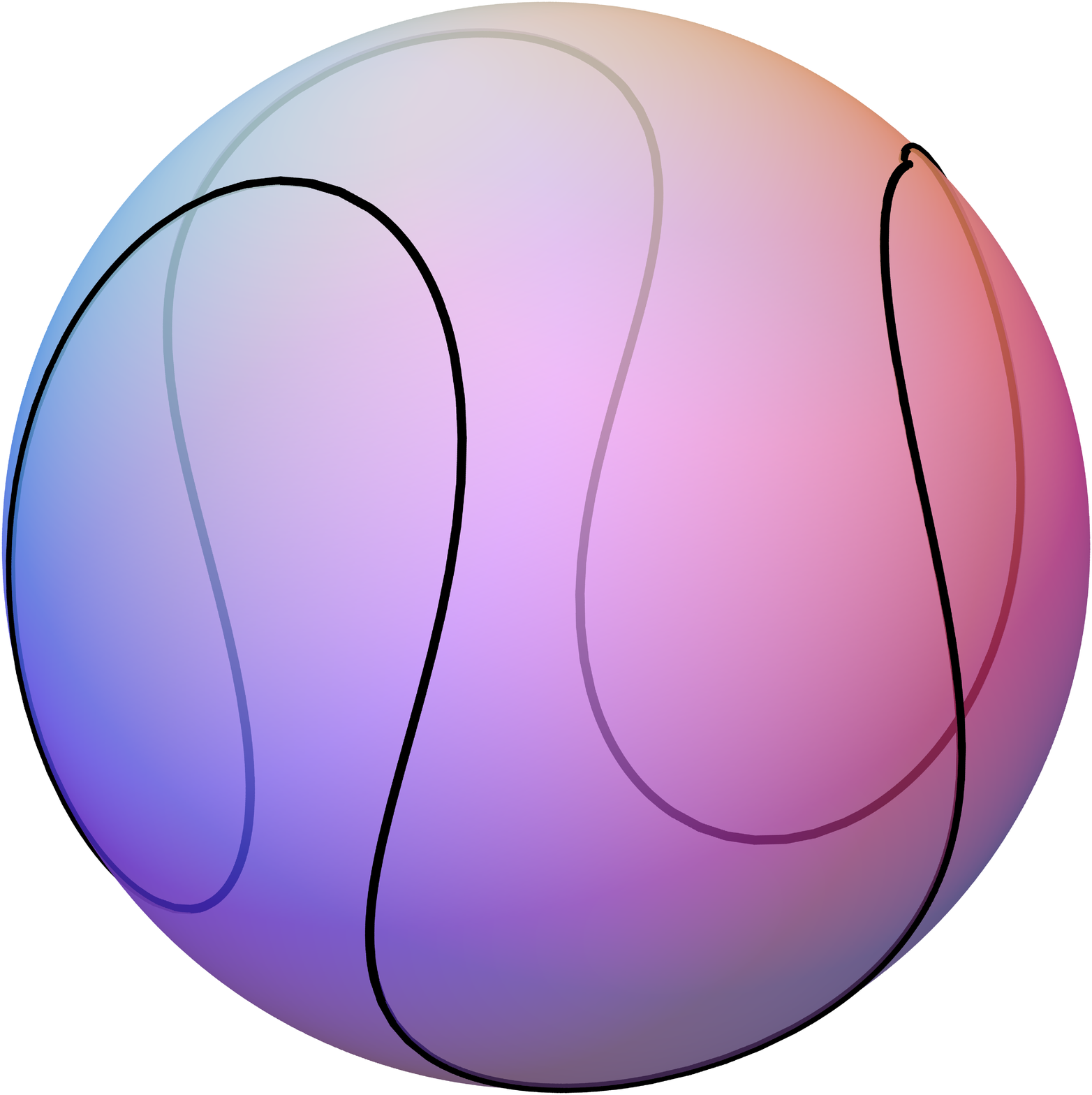}
\includegraphics[width=1.4in,angle=80]{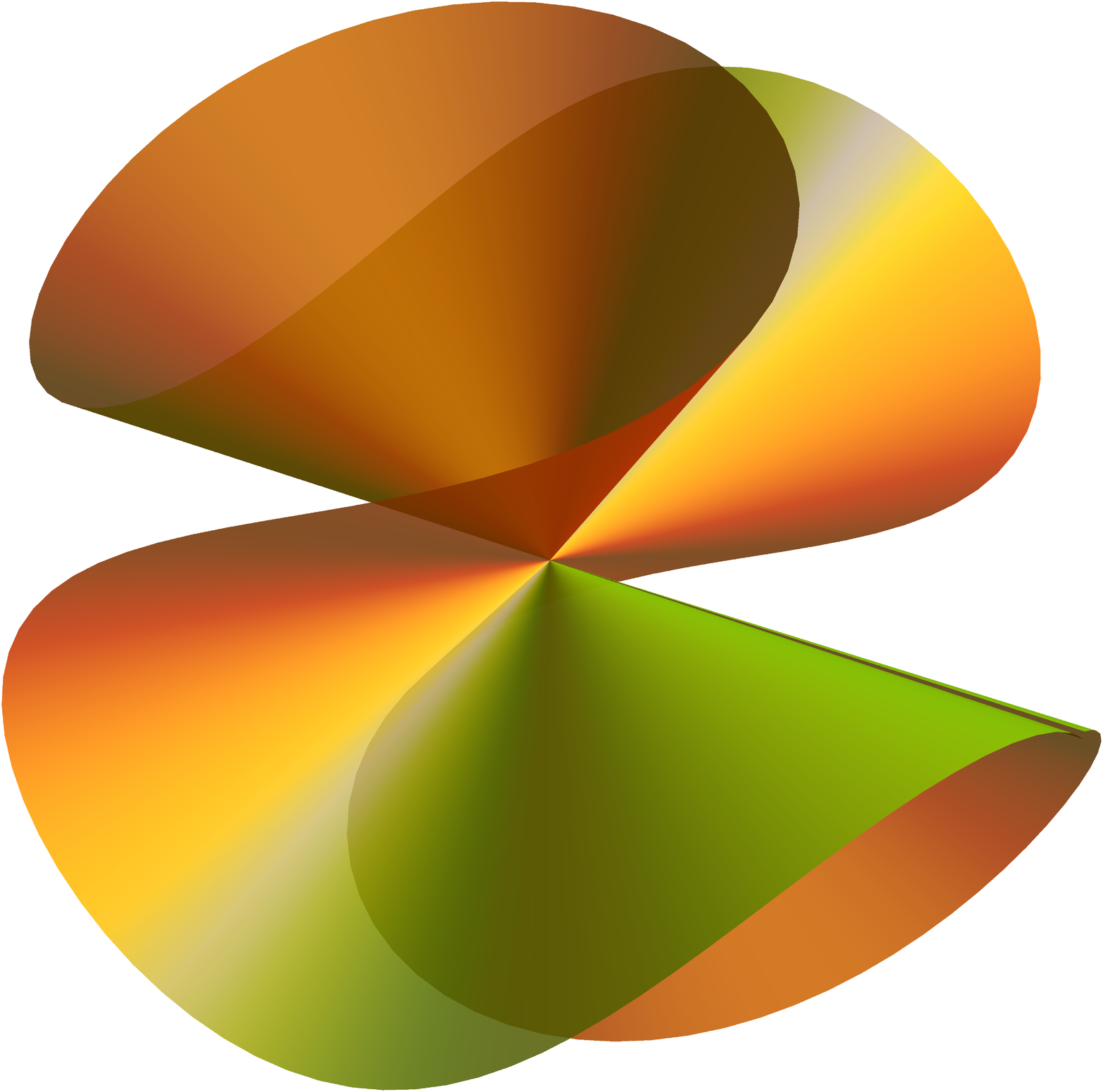}

\caption{(Color online) (Left) The curve $\ugamma(s)$ on the surface of a (unit) sphere.  (Right) The ruffle surface generated by families of the the curves. A measure of the extra length accommodated by the ruffle is ratio of the (new) perimeter $L$ to the great circle length $2\pi$, here $L/2\pi = \lambda^{1+ \nu} = 2.20$, and $n=3$.}\label{fig:ruffles}
\end{figure}

The essence of the normal bend energy of the anti-cone, that is the $\left((\hat{\rvec}\wedge \ugamma')\cdot \ugamma''\right)^2$ factor in eq.~(\ref{eq:normalbend_energy}) associated with the generator curves, can be re-written in terms of the Darboux frame with respect to the surface of the sphere of radius $r=1$ on which the curve $\ugamma(s)$ can be also thought to live:  $\vec{T}(s) = \ugamma'(s)$ as before,  and $\hat{\rvec} = \vec{u}\s{s}(s)$ where the subscript $\s{s}$ denotes ``on the sphere".  The third member of the triad is the curve's tangent normal on the sphere $\vec{t}\s{s}(s) = \vec{u}\s{s}\wedge\vec{T} = \hat{\rvec}\wedge \ugamma' = \vec{u}$.  Thus the roles between the two frames have been interchanged: $\vec{u},\vec{t} \rightarrow \vec{t}\s{s}, \vec{u}\s{s}$.  The normal curvature, eq.~(\ref{eq:normalbend}), on the anti-cone can be re-written as $\vec{t}\s{s} \cdot \vec{T}' =\kappa\s{g}\sp{s} $, that is, it is the geodesic curvature of the generator curve, but on the sphere (here $\sp{s}$ also denotes ``on the sphere").  The normal bend energy of the anti-cone that we require to minimize is the same as minimizing $\int\d A \kappa\s{g}\sp{s 2} $, which is the problem of minimizing the geodesic curvature of an elastica on a sphere, see Langer and Singer \cite{Singer:84,Singer:87,Singer:08,Singer:08b}, which we now employ:

The curve on a sphere with minimal geodesic curvature energy has curvature as a function of arc length $s$ of:
\be
\kappa\s{g}\sp{s} = \kappa_0 {\rm cn}(\frac{\kappa_0 s}{2p},p)\label{eq:minimal_bend}
\ee
where ${\rm cn}$ is a Jacobi elliptic function, and $\kappa_0$ is the maximal curvature along the curve. The curve must be periodic in the arc length, that is $\frac{\kappa_0}{2p} L = n 4 {\rm K}(p) $
with ${\rm K}(p)$ being the complete elliptic integral of the first kind, since $4{\rm K}(p)$ is the period of ${\rm cn}$, and $L=2\pi \lambda^{1+\nu}$ is the (new) perimeter in terms of the (new) radius $r=1$. As before, $n$ is the number of cycles in one revolution.  The parameter $p$ ensures that the curve closes up on the sphere's surface -- $p= 0.51$ in the illustration.  Thus the maximum curvature is fixed by
\be
\frac{\kappa_0}{p} \pi \lambda^{1+\nu} = n 4 {\rm K}(p) \label{eq:length}.
\ee
 Since $\vec{T}'^2 = \ugamma''^2 = \kappa\s{g}\sp{s 2}  + \kappa\s{n}\sp{s 2}  = \kappa\s{g}\sp{s 2}  + \frac{1}{r^2} = \kappa\s{g}\sp{s 2}  + 1$, to plot in real space one needs to solve $\ugamma''^2 = 1 + \kappa\s{g}\sp{s 2} $, with $\kappa\s{g}\sp{s}$ given by Eq.~(\ref{eq:minimal_bend}) for the trajectory $\ugamma(s)$.  Results are of the form of Fig.~\ref{fig:ruffles}(left) for the trajectory $\ugamma(s)$ on a sphere, and of Fig.~\ref{fig:ruffles}(right) for the surface generated by it.
Note in particular that the inflection points of in $\kappa\s{g}\sp{s}$ are on a closed geodesic (here the equatorial great circle), there is re-entrance, there is no constraint of $\phi = \Phi$, and, as expected by convexity, there are no sharply bent ridges where the bend energy would is large.  For extreme elongations (large $\lambda$ as found in elastomers) there is so much extra arc length that has to be stored in the convoluted, bent trajectory that the curves eventually intersect themselves, forming orbit-like solutions \cite{Singer:84,Singer:87,Singer:08,Singer:08b} that are clearly not accessible to our deforming, impenetrable surfaces, which would then deviate from the solutions of Langer and Singer.

Note that the small amplitude form adopted, Eq.~(\ref{eq:hdef}), represents the appropriate limiting case of this general solution. For small amplitudes, the curves generating the anti-cone never deviate far from the equator of the sphere. Gauss curvature is a second order effect in distance and so the curve can now be thought of in this limit as lying instead across a cylinder. The problem is that of periodic elastica in the plane with harmonic forms for displacement and curvature. To explicitly demonstrate that the form (\ref{eq:hdef}) is the minimizer, first note that the LS minimal curvature (\ref{eq:minimal_bend}) for small differences in length between the perimeter and the great circle, $ p \rightarrow 0$, is $\kappa\s{g}\sp{s} \rightarrow  \kappa_0 \cos(qs)$, taking the limit of the elliptic function for small modulus, and where maximal curvature also becomes small with $p$ such that $q =\kappa_0 /2p$ is finite.
The general form (\ref{eq:normalbend}) of the normal curvature can be explicitly calculated for trajectories $\ugamma(s) = (\sin\theta\sin\phi, \sin\theta\cos\phi,\cos\theta)$, in usual spherical coordinates, in terms of $\phi(s)$, $\theta(s)$ and their derivatives.  From the harmonic form $h(\phi)$, Eq.~(\ref{eq:hdef}), we showed the azimuthal variation $\d \phi/\d s$ is given by Eq.~(\ref{eq:new-dp}), and one can easily find $\tan\theta(\phi) = 1/(A\sin(n\phi))$ and thence $\d \theta/\d \phi = -nA\cos(n\phi)/\left(1+ A^2\sin^2(n\phi)\right)$ and thus also derivatives of $\theta(s)$.  Injecting these relations into the curvature, keeping only terms of order $A$, one obtains  $\kappa\s{g}\sp{s} = A(n^2-1)\sin(n\phi)$, sine rather than cosine appearing because of a choice of phase.

Returning to Eq.~(\ref{eq:phi-length}) but integrating up to a $\phi$ and an $s$ rather than a complete revolution, one obtains
\be
 \left[1 + \frac{A^2}{4} (n^2 -1) \right]\phi + \dots  = s\nonumber .
\ee
Performing the complete integral, having expanded to lowest order in $A^2$, gives an explicit form for the amplitude:
\be 
A = 2 \sqrt{\frac{\lambda^{1+\nu} - 1}{n^2 - 1}}\label{eq:small-A}.
\ee 
Taking these $\phi(s)$ and $A(\lambda)$ in the curvature above then gives the LS, minimal form $\kappa\s{g}\sp{s}  = \kappa_0 \cos(qs)$, showing that the $h(\phi)$ in Eq.~(\ref{eq:hdef}) is the bend minimizing harmonic form.

By identification, one obtains the curvature amplitude and the wave-vector as functions of extension $\lambda$:
\be
\kappa_0 =  2 \sqrt{(\lambda^{1+\nu} - 1)(n^2 - 1)} \;\;\;\;\;\;\;\;\; q = n/\lambda^{1+\nu} \label{eq:max-min-curvature-q-vector}.
\ee
One must recall that we are dealing with a unit sphere in the target space, which has set the scale for $s$ and $q$.

Elastomers were not discussed in \cite{ModesPRE:10,ModesPRS:11} -- they are more subtle than glasses since they can sometimes alleviate stress by director rotation.  For instance in the azimuthal example considered above and by Pismen, if an \textit{elastomer} disc were held flat on cooling so that extensile radial stress developed because of the deficit of natural length in the radial direction, then director rotation from the azimuthal towards the radial direction would re-attribute length from the azimuthal to radial direction, i.e. from a direction of (azimuthal) surplus to one of (radial) deficit.  Little or no energy cost for such a distortion is required -- so-called soft elasticity \cite{soft1}  -- and the need for anti-cones obviated.  Thus general analyses of nematic elastica mentioning nematic elastomers require caution.

An elastomer example that \textit{would} produce anti-cones would be a radial, 2-D, +1 defect being heated, that is $\lambda <1$; in Fig.~\ref{fig:anticones}(a) there would instead be radial lines of $\n$ in the initial, undistorted disc shown in the equatorial plane.  Now a circular path obtains surplus length (by a factor of $\lambda^{-\nu}$) and a radius is in deficit (by a factor of $\lambda$).  The radius cannot become longer by rotation of the director towards it -- it is already radial, and the glass and elastomer responses are identical in character.  The difference is that elastomer contractions can be huge, for instance $\lambda \rightarrow 0.25$ on heating ($\rightarrow 4$ on cooling) is possible, and the topography changes could be accordingly larger than in glasses.

We have explored and contrasted the differing assumptions one can make about deformations involved in anti-cone formation, principally the role of azimuthal displacements of material points.  Bend energy-minimising configurations have been explicitly demonstrated.  They embrace large deformations and re-entrance.  We also point to possible experiments, on elastomers rather than glasses, where effects are very large.

\bibliography{References_disclinations}

\end{document}